# Strain-tuning Bloch- and Néel-type magnetic skyrmions: a phase-field simulation

Shouzhe Dong[a,b], Jing Wang[a], Xiaoming Shi[a], Deshan Liang[a], Hasnain Mehdi Jafri[a], Chengchao Hu[c], Ke Jin [a]*, Houbing Huang[a]*

[a]Advanced Research Institute of Multidisciplinary Science, Beijing Institute of Technology, Beijing, 100081, China

[b]School of Mechatronic Engineering, Beijing Institute of Technology, Beijing, 100081, China

[c]College of Materials Science and Engineering, Liaocheng University, Liaocheng, 252059, China

## Abstract

Strain manipulation of the magnetic domains, such as the stripe domains and skyrmions, has attracted considerable attention because of its potential applications for magnetic logic and memory devices. Here, utilizing phase-field modeling, we demonstrate the deterministic modulation of the orientation and the configuration of the stripe domains and skyrmions by using a uniaxial strain. The reorientation of the stripe domains can be caused by a suitable strain, and the direction of the reorientated domains is determined by the direction of the applied uniaxial strain and the type of domain walls, including Bloch- and Néel- types. Furthermore, by constructing a phase diagram, we discovered that when the uniaxial tensile strain increases, the ferromagnetic islands undergo a continuous phase transition from a skyrmion to multi-domains or a single domain. The competition between magnetic anisotropy energy and stray field energy leads to the continuous phase transition and the formation of domain patterns under the uniaxial tensile strain. Our research provides a theoretical foundation for the development of strain-controlled magnetic domain designs.

*Corresponding author: hbhuang@bit.edu.cn (Houbing Huang), jinke@bit.edu.cn (Ke Jin)

Domain walls [1–3], with the transition area separating the two domains with different orientations, are still one of the central focuses of modern material science and physics. Based on the transition type between magnetic domains, the one-dimensional spin chain may be divided into Bloch-type spin chains (see Fig. 1(a)) and Néel-type spin chains (see Fig. 1(b)). Similarly, the two-dimensional 180° magnetic domain walls can be divided into Bloch-type domain walls (see Fig. 1(c)) and Néel-type domain walls (see Fig. 1(d)). In addition, skyrmion [4–8] as a nanoscale topological spin texture, can also be classified as Bloch-type skyrmion (see Fig. 1(e)) and Néel-type skyrmion (see Fig. 1(f)) according to the transition way of the spins from inside to outside position. These rich spin textures [9,10] result from the mutual competition among stray field energy, magnetocrystalline anisotropy energy, exchange energy, Dzyaloshinskii-Moriya interaction (DMI) energy, and elastic energy. In addition, the magnetic domain structures with different magnetic domain wall types also have different energy competition modes. Therefore, it is crucial to take into account how the types of magnetic domain walls affect the evolution of the magnetic domain structure under the excitation of an external field.

Concurrently, these low-dimensional spin textures can be created and manipulated by magnetic field [10,11], spin-polarized current [13, 14], spin waves [15, 16], or other methods [16–18], implying potential applications in energetically efficient magnetic memories, logic gates, synaptic devices, *etc* [9, 20]. However, the above techniques show inevitable disadvantages such as low integration density and high Joule heating effect. In recent years, magnetoelectric coupling provides a new perspective for us to control magnetic domains [20–22], and this method can overcome these drawbacks. One of the methods is to select the ferroelectric thin film as the substrate, then a large in-plane strain is generated during the reversal of ferroelectric polarization in the ferroelectric layer, which is expected to significantly modify the interfacial magnetism of the ferromagnetic layer [23,24]. Thus, a reliable and controllable transition between skyrmions and other magnetic states may be envisioned. Even though much study has been done on the use of stress or strains to control domain structures and their distribution [25–30], there is still a lack of understanding of the mechanism for strain-tuning domains with different types of domain walls.

In the present work, we investigate the uniaxial strain modulation of magnetic orders in ferromagnetic thin films and islands with different types of domain walls by a phase-field model. Different from the previous study on strain-induced magnetic anisotropy variation [25, 31, 32], a magnetoelastic coupling of magnetization and strain is introduced in the phase-field model to describe the anisotropic deformation of magnetic domain lattices. By constructing two models with different domain wall types, it is demonstrated that uniaxial tensile strain in ferromagnetic thin films or islands with different types of domain walls can induce a different variation of magnetic domain structures. The results of our research have great theoretical guiding significance for experimentally designing certain strain-tuning magnetic domain configuration devices.

In the phase-field model [33–37], the spatial distribution of the magnetization field $\mathbf{M} = M_s\mathbf{m} = M_s(m_1, m_2, m_3)$ is used to describe the domain structure, where $M_s$ and $\mathbf{m}$ are the saturation magnetization and the components of unit magnetization vector, respectively. The transient magnetization domain structure is described by the Landau-Lifshitz-Gilbert (LLG) equation:

$$(1+\alpha^2)\frac{\partial \mathbf{M}}{\partial t} = -\gamma_0 \mathbf{M} \times \mathbf{H}_{eff} - \frac{\gamma_0 \alpha}{M_s}\mathbf{M} \times (\mathbf{M} \times \mathbf{H}_{eff}) \quad (1)$$

where $\alpha$ is the damping constant, $\gamma_0$ is the gyromagnetic ratio, and $\mathbf{H}_{eff}$ is the effective magnetic field, which can be given as

$$\mathbf{H}_{eff} = \frac{-1}{\mu_0 M_s}\frac{\partial F_{tot}}{\partial m} \quad (2)$$

where $\mu_0$ and $F_{tot}$ denote vacuum permeability and total free energy, respectively. The total free energy can be written as

$$F_{tot} = F_{stray} + F_{anis} + F_{exch} + F_{dmi} + F_{elas} \quad (3)$$

where $F_{stray}$, $F_{anis}$, $F_{exch}$, $F_{DMI}$, and $F_{elas}$ are the stray field energy, magnetocrystalline anisotropy energy, exchange energy, DMI energy, and elastic energy, respectively. The magnetostatic energy of a system can be written as

$$F_{stray} = -\frac{1}{2}\mu_0 M_s \iiint \mathbf{H}_d \cdot \mathbf{m}\, dV \tag{4}$$

where $\mathbf{H}_d$ is the demagnetization field, determined by the long-range interaction among the magnetic moments in the system, which is governed by

$$\mathbf{H}_d = H_{d1,1} + H_{d2,2} + H_{d3,3} = -M_s(m_{1,1} + m_{2,2} + m_{3,3}) \tag{5}$$

Two types of magnetocrystalline anisotropy are considered cubic anisotropy with magnetic easy axes along with <100>, <110>, or <111> crystal axes, and the uniaxial anisotropy in ultra-thin films with a magnetic easy/hard axis perpendicular to the film plane. The volume of magnetocrystalline anisotropy energy is given by:

cubic anisotropy

$$F_{anis} = \iiint_V [K_1(m_1^2 m_2^2 + m_2^2 m_3^2 + m_3^2 m_1^2) + K_2\, m_1^2 m_2^2 m_3^2]\, dV \tag{6}$$

and uniaxial anisotropy

$$F_{anis} = \iiint_V [K_1(1 - m_3^2) + K_2(1 - m_3^2)^2]\, dV \tag{7}$$

where $K_1$ and $K_2$ are the anisotropy constants. The exchange energy is determined by the spatial variation of the magnetization orientation and can be described as

$$F_{ex} = A\int_V m_{1,1}^2 + m_{1,2}^2 + m_{1,3}^2 + m_{2,1}^2 + m_{2,2}^2 + m_{2,3}^2 + m_{3,1}^2 + m_{3,2}^2 + m_{3,3}^2\, dV \tag{8}$$

where $A$ is the exchange stiffness constant. The comma in the subscript denotes spatial differentiation. For instance, $m_{i,j} = \dfrac{\partial m_i}{\partial x_j}$, where $x_j$ is the $j$th component of the position vector in the Cartesian coordinates.

Two types of DMI are considered, one interaction can be invoked when simulating an ultrathin magnetic thin film or island with interface spin-orbit coupling. Considering a homogenous effective DMI constant $D_{\text{interface}}$, the interface DMI energy is given by

$$F_{dmi} = \iiint_V D_{\text{interface}} (m_3 \nabla \cdot m - m \cdot \nabla m_3) dV \tag{9}$$

Another kind of DMI can be invoked when simulating some cubic materials whose spatial symmetry is broken. Also, considering a homogenous effective DMI constant $D_{\text{bulk}}$, the bulk DMI energy is given by

$$F_{dmi} = \iiint_V 2 D_{\text{bulk}} (m \cdot \nabla \times m) dV \tag{10}$$

The elastic energy generated from the local deformation can be described as

$$F_{el} = \iiint_V \frac{1}{2} c_{ijkl} e_{ij} e_{kl} dV = \int_V \frac{1}{2} c_{ijkl} (\varepsilon_{ij} - \varepsilon_{ij}^0)(\varepsilon_{kl} - \varepsilon_{kl}^0) dV \tag{11}$$

where $c_{ijkl}$, $e_{ij}$, $\varepsilon_{ij}$ and $\varepsilon^0_{ij}$ are the elastic stiffness tensor, elastic strain, total strain, and stress-free strain ($i, j, k, l$ = 1, 2, 3, 4). For a cubic magnetostrictive material, the local magnetization can be represented by the stress-free strain as following

$$\varepsilon_{ij}^0 = \begin{cases} \frac{3}{2} \lambda_{100} (m_i m_j - \frac{1}{3}) & (i = j) \\ \frac{3}{2} \lambda_{111} m_i m_j & (i \neq j) \end{cases} \tag{12}$$

where $\lambda_{100}$ and $\lambda_{111}$ are magnetostrictive constants. With three independent elastic constants $c_{11}$, $c_{12}$, and $c_{44}$ in Voigt's notation, the elastic energy can be rewritten as

$$F_{el} = \iiint_V [\frac{1}{2}c_{11}(e_{11}^2 + e_{22}^2 + e_{33}^2) + c_{12}(e_{11}e_{22} + e_{22}e_{33} + e_{11}e_{33}) \\ + 2c_{44}(e_{12}^2 + e_{23}^2 + e_{13}^2)]dV$$

$$= \iiint_V \begin{cases} \frac{1}{2}c_{11}[(\varepsilon_{11} - \varepsilon_{11}^0)^2 + (\varepsilon_{22} - \varepsilon_{22}^0)^2 + (\varepsilon_{33} - \varepsilon_{33}^0)^2] \\ + c_{12}[(\varepsilon_{11} - \varepsilon_{11}^0)(\varepsilon_{22} - \varepsilon_{22}^0) + (\varepsilon_{22} - \varepsilon_{22}^0)(\varepsilon_{33} - \varepsilon_{33}^0) \\ + (\varepsilon_{11} - \varepsilon_{11}^0)(\varepsilon_{33} - \varepsilon_{33}^0)] + 2c_{44}[(\varepsilon_{12} - \varepsilon_{12}^0)^2 + (\varepsilon_{23} - \varepsilon_{23}^0)^2 \\ + (\varepsilon_{13} - \varepsilon_{13}^0)^2] \end{cases} dV \qquad (13)$$

Homogeneous strain denotes the macroscopic shape variation by the domain structure formation, while the heterogeneous strain does not affect the macroscopic shape.

The simulation in this work uses FeGa alloy and $Co_{20}Fe_{60}B_{20}$ as model systems to study the strain effect on the magnetic domains with different domain walls types. The materials parameters for ultrathin $Co_{20}Fe_{60}B_{20}$ used in simulations are listed as follows [24, 35–38]: $A_{ex}$ = 1.9 × 10$^{-11}$ J/m, $K_1$ = 9.78 × 10$^5$ J/m$^3$, $K_2$ = 1.108 × 10$^4$ J/m$^3$, $M_s$ = 1.25 × 10$^6$ A/m, $D_{interface}$ = 0.75 mJ/m$^2$; $\gamma_0$ = 2.2 × 10$^5$ mA$^{-1}$·s$^{-1}$, $\lambda_{100} = \lambda_{111} = \lambda_s$ = 3.7 × 10$^{-5}$, $c_{11}$= 218.1 GPa, $c_{12}$= 93.46 GPa; and the parameters for FeGa employed in the simulations are as follows [29,42]: $A_{ex}$ = 2 × 10$^{-11}$ J/m, $K_1$ = 2 × 10$^4$ J/m$^3$, $K_2$ =0 J/m$^3$, $M_s$ = 1.432 × 10$^6$ A/m, $D_{bulk}$ = 5.03 mJ/m$^2$; $\gamma_0$ = 2.2 × 10$^5$ mA$^{-1}$·s$^{-1}$, $\lambda_{100}$ = 2.64 × 10$^{-4}$, $\lambda_{111}$ = 0, $c_{11}$ = 196 GPa, $c_{12}$ = 156 GPa. In the present simulations, the elastic stiffness tensor for the ferroelectric substrate is set as the same as that of the magnetic island, i.e., $c_p$ = $c_{island}$ for simplicity.

The film size is 500 × 500 × 425 nm$^3$, and the number of cells is 80 × 80 × 17. Specifically, we set up 17-layer grids, of which the bottom 11-layer grid describes the elastic rigid piezoelectric layer, the middle 2 layers describe the magnetic ultrathin film, and the top 4 layers describe the space on top of the magnet [23]. Without applying an external magnetic field and strain, the magnetic ground states of FeGa and $Co_{20}Fe_{60}B_{20}$ films are the labyrinth domain, and the particular magnetic domain distribution is shown in Fig. 2 (a) and (d), respectively. The difference is that the DMI in FeGa arises from the breaking of its spatial inversion symmetry, which results in a helical phase that belongs to the striped domain with the Bloch-type domain walls; whereas the DMI in $Co_{20}Fe_{60}B_{20}$ is introduced by interface spin coupling, which origins from the interface between the heavy metal and the magnetic layer. Because of the competition between exchange coupling energy and DMI energy, it tends to produce a striped domain with a Néel-type domain wall transition mode.

First, we explored the evolution of striped domains with Bloch-type domain walls under the regulation of uniaxial tensile strain. As shown in Fig. 2(b), compared to the unstrained labyrinth-type striped domain, a 0.5% uniaxial tensile strain in the y-direction can align the domain walls orientation of the stripe domain to the uniaxial tensile strain direction (the strain distribution can be seen in Fig. S1 in supplementary materials). When we apply a uniaxial compressive strain in the y-direction, the domain wall's orientation can finally be perpendicular to the case by applying a uniaxial tensile strain, as shown in Fig. 2(c). This is also consistent with previous reports [29].

On the other hand, the evolution of the striped domains with Néel domain walls under the application of uniaxial tensile strain is opposite to the case of Bloch domain walls. When a 0.5% uniaxial tensile strain is applied in the y-direction, the labyrinth domain evolves into a very standard striped domain, with the direction of the domain walls perpendicular to the direction of tensile strain, but when a -0.5% uniaxial compressive strain is applied in the y-direction, the labyrinth domain evolves into a straight striped domain, with the direction of the domain walls is consistent with the direction of strain, as shown in Fig. 2(e) and Fig. 2(f).

Furthermore, what effect would uniaxial tension strain have on the topologically protected magnetic structure? As a consequence, we constructed two types of nano-islands that use distinct DMI energy forms to maintain Bloch- and Néel-type skyrmions. The island size is $236 \times 236 \times 9.35$ nm$^3$, and the number of cells is $118 \times 118 \times 17$. Specifically, we set the 17-layer grids, of which the bottom 11 layers of grids describe the elastically stiff piezoelectric layer, the middle 2 layers describe the magnetic island and the top 4 layers describe the space on top of the magnet. The diameter of the magnetic island is 220 nm. The parameters we choose to construct Néel here are the same as Hu's work [24]. For comparison, Bloch skyrmion is constructed and also uses these parameters, only the interface DMI is replaced by the Bulk DMI of the material.

Fig. 3(a) and (d) show the model diagrams of the Bloch skyrmion and the Néel skyrmion stably exists on the ferroelectric substrates, respectively. The strain effect on magnetic skyrmion is studied by modulation of the ferroelectric polarization. The Bloch skyrmion on the nano-island can maintain a nearly perfect circle when zero strain is applied, as shown in Fig. 3(b). But Fig. 3(c) shows that when we apply 0.5% tensile strain in the y-direction, the circular skyrmion will deform and turn into an elongated skyrmion, with the long axis parallel to the direction of uniaxial strain. This skyrmion distortion shows the similar result with the case of tensile strain in the y-direction on the Bloch-type maze domain. For the Néel skyrmion, a 0.5% uniaxial tensile strain applied in the y-direction can also make the circular skyrmion deforms into an oval skyrmion, but the long axis of the ellipse is perpendicular to the uniaxial strain direction.

Similar to how uniaxial strain modulates the orientation of the striped domains with the two domain wall types described above, the variation trend of the skyrmions with the two domain wall types is the same. By analyzing the energy behavior in the magnetic domain variation (see Fig. S2 in supplementary materials), their ultimate goal is to promote the consistency of the magnetic moment orientation and the external strain orientation, as well as to meet the goal of minimum energy. Specifically, the equation of the magnetoelastic energy can be expressed as $E_\sigma = -(2/3)\lambda\sigma\cos^2\theta$, where $\lambda$ is the magnetostriction coefficient, $\sigma$ is the stress and $\theta$ is the angle between the direction of strain and the orientation of spontaneous magnetization. It can be seen from this formula that when $\lambda>0$, $E_\sigma$ will be minimized when $\theta$ is 0° or 180° under tensile stress ($\sigma > 0$). Because of the different transition modes of Bloch- and Néel-type domain walls, the direction of domain walls keeps a parallel or perpendicular distribution relationship with the direction of the strain. We demonstrate the essence of uniaxial strain-tunning the domain structures with different types of magnetic domain walls, which provided theoretical direction for more accurate experimental analysis and manipulation of the magnetic domain structures.

To identify the stable magnetic orders under different DMI constants and uniaxial strains, Fig. 4 presents the strain-DMI-phase diagram of Néel skyrmion and Bloch skyrmion based on phase-field simulation. Fig. 4(a) shows the stable magnetic structure of FeGa magnetic islands with Bloch skyrmion as the initial magnetic structure under different uniaxial strains and DMI coefficients. It is found that there are two different phases, namely the multidomain and skyrmions phases, in the islands when subjected to different strains and DMI constants. When the strain is less than 0.8%, the DMI dominates and the island always has a skyrmion magnetic domain. As the uniaxial strain increases, the circular skyrmion becomes distorted and gradually transforms into an elliptical skyrmion, with the long axis direction of the ellipse skyrmion aligning along the strain direction. However, when the strain exceeds 0.8%, the strain takes over and the skyrmion begins to annihilate, resulting in a multi-domain state. This result arises from the system's elasticity being minimized. Compared with Bloch-type skyrmion, Néel skyrmion has to go through three processes: skyrmion-to-multidomain phase, and multidomain-to-monodomain phase. When there is no applied strain and the DMI constant is 0.75 mJ/m$^2$, skyrmion can be stable. As the tensile strain increases, the skyrmion begins to distort into an elliptical skyrmion, with the primary axis of the ellipse skyrmion perpendicular to the direction of uniaxial strain due to the effect of elastic energy. The skyrmion was destroyed and the system transitioned to a monodomain state to minimize the system's energy when the strain was larger than 0.8% [30]. However, when the DMI constant is more than 1 mJ/m$^2$, a multidomain state exists between the skyrmion phase and the monodomain phase. The existence of the multidomain state should be explained by the fact that increasing the DMI action induces an increase in magnetic moment nonlinearly. By establishing a phase diagram, it is possible to guide experimental guys to better design skyrmion-based storage devices.

In short, we investigate the uniaxial strain modulation of magnetic orders in magnetic thin films and islands with different domain wall types by a phase-field model. Using an appropriate uniaxial tensile strain, the direction of the Bloch-type domain walls can be modulated consistent with the tensile strain direction, but using an appropriate uniaxial tensile strain, the direction of the Néel-type domain walls is perpendicular to the direction of strain. The strain causes two skyrmions to deform in opposing directions, which is similar to how uniaxial strain affects stripe domains. We demonstrate that this domain pattern variation results from the competition between magnetic anisotropy energy and stray field energy through the analysis of energy density change and domain structure. Finally, due to the influence between elastic and DMI energy, we explored the impacts of DMI constants and uniaxial strain on the two types of skyrmions, and the phase diagram indicates a complex structure with skyrmions, multi-domains, and monodomains. Our results provide the guidance on manipulating magnetic domain structure for the experiments. And uniaxial strain tunning of distinct magnetic phases in magnetic thin films and islands opens the path for future spintronic and memory devices.

**Declaration of Competing Interest**

The authors declare that they have no known competing financial interests or personal relationships that could have appeared to influence the work reported in this paper.

**Acknowledgment**

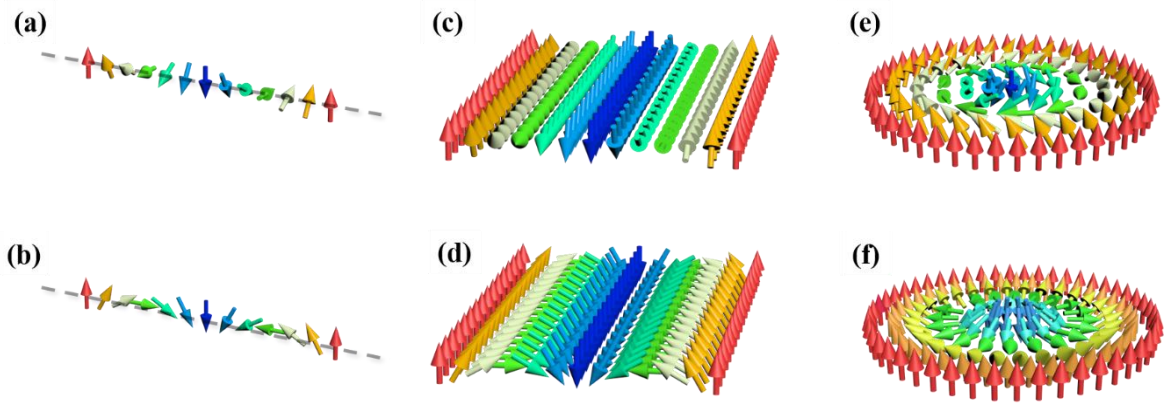

**Fig. 1.** Spin structures with a different type of spin arrangement. (a), (c), (e) Bloch type spin chain, strip, and skyrmion, respectively; (b), (d), (f) Néel type chain, strip, and skyrmion, respectively.

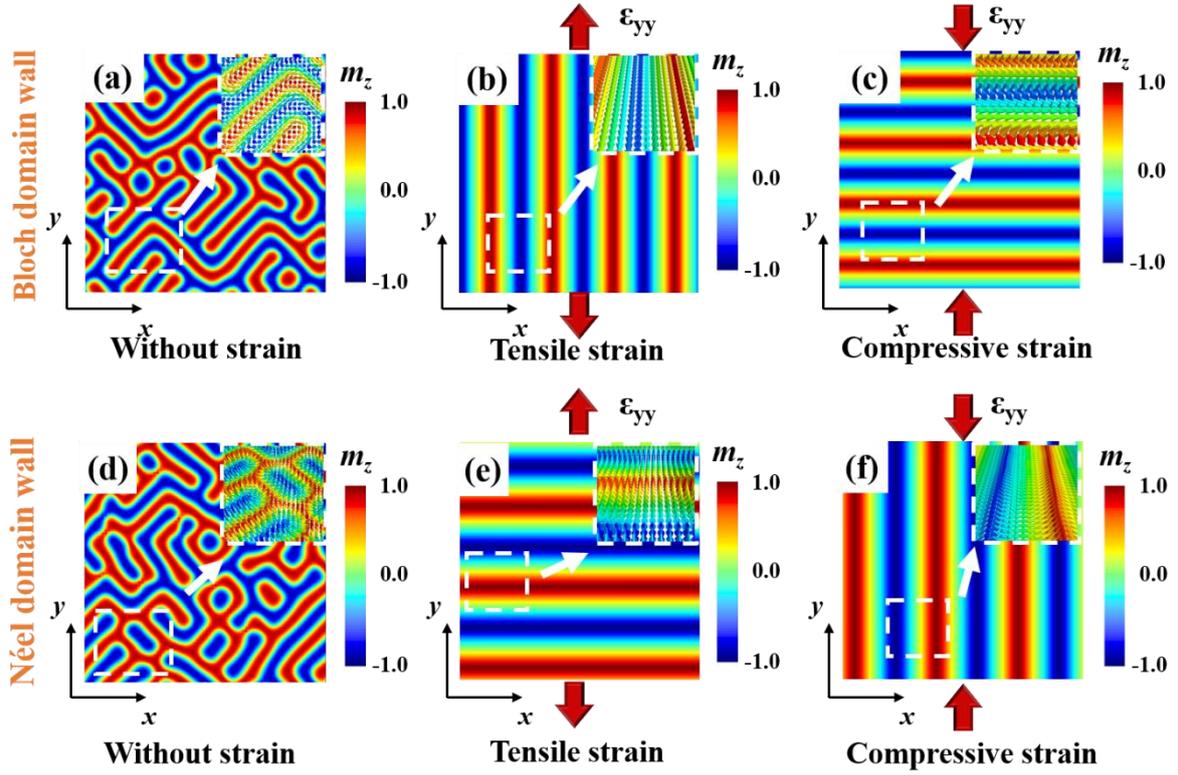

**Fig. 2.** (a)-(c) The evolution of Bloch type labyrinth domain under (a) zero strain, (b) 0.5 % tensile strain and (c) -0.5 % compressive strain in the y-direction. (d)-(f) The evolution of Néel-type labyrinth domains under (d) zero strain, (e) 0.5 % tensile strain and (f) -0.5 % compressive strain in the y-direction. The inserts show the enlarged view of the magnetic moment in the white dashed box, which identifies the type of domain walls.

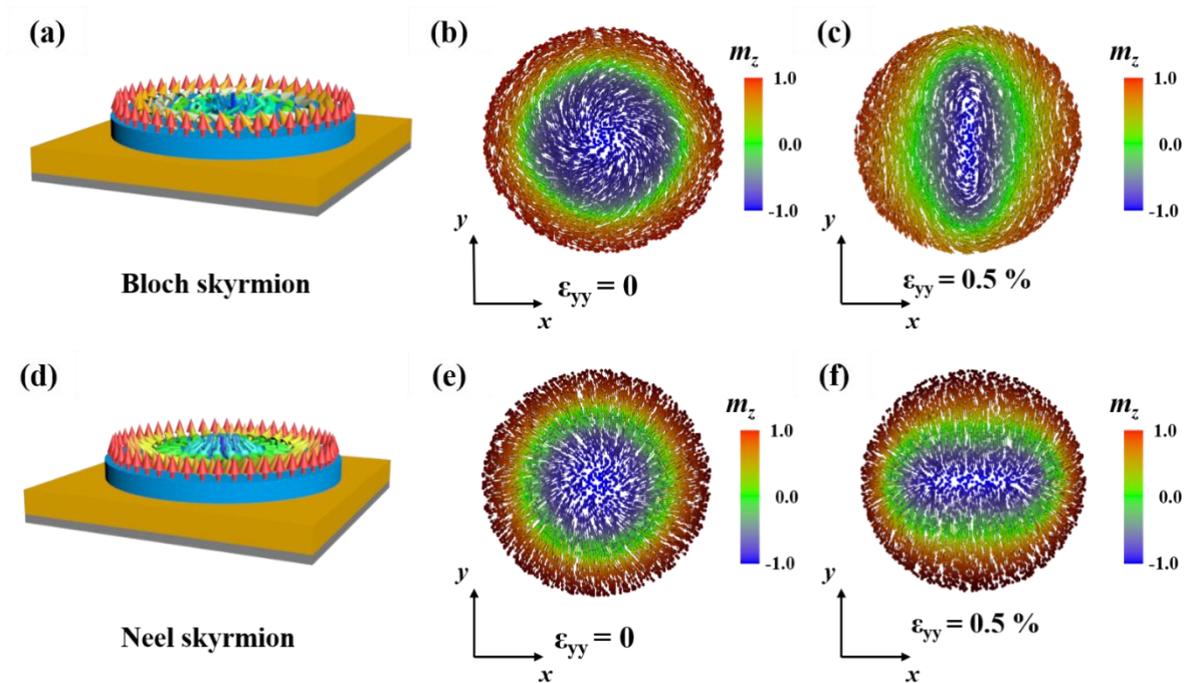

**Fig. 3.** (a) Schematic diagram of the Bloch skyrmion on the nanoisland; The evolution of Bloch skyrmion under (b) zero strain and (c) 0.5 % tensile strain in the y-direction; (d) Schematic diagram of the Néel skyrmion on the nanoisland; The evolution of Néel skyrmion under (e) zero strain and (f) 0.5 % tensile strain in the y-direction.

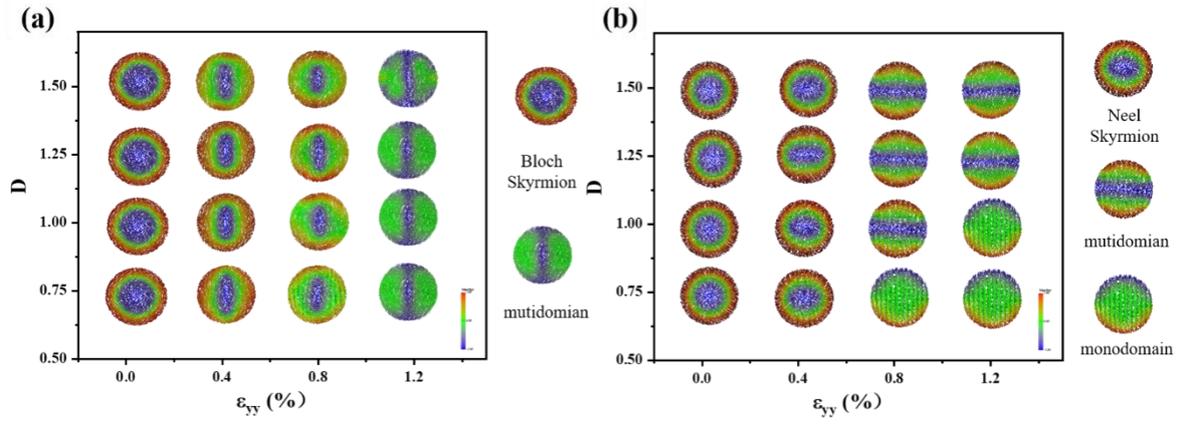

**Fig. 4.** Phase diagram for Bloch skyrmion (a) and Néel skyrmion (b) in terms of uniaxial strain $\varepsilon_{yy}$ versus DMI constant.

# Strain-tuning Bloch- and Néel-type magnetic skyrmions: a phase-field simulation

Shouzhe Dong[a,b], Jing Wang[a], Xiaoming Shi[a], Deshan Liang[a], Hasnain Mehdi Jafri[a], Chengchao Hu[c], Ke Jin[a*], Houbing Huang[a*]

[a]Advanced Research Institute of Multidisciplinary Science, Beijing Institute of Technology, Beijing, 100081, China

[b]School of Mechatronic Engineering, Beijing Institute of Technology, Beijing, 100081, China

[c]College of Materials Science and Engineering, Liaocheng University, Liaocheng, 252059, China

*Corresponding author: hbhuang@bit.edu.cn (Houbing Huang), jinke@bit.edu.cn (Ke Jin)

**Supplementary Note 1: Simulated strain distribution of the phase-field model.**

In general, there will always be a strain gradient when the thin film is grown on the substrate [1–4]. In our simulation, we also considered the strain gradient both for the thin film and the nano-island. Fig. S1(b) shows the strain distribution in the magnetic island with a size of 220 nm in our model when a 0.5% uniaxial tensile strain is applied. It can be noticed that some strain gradients exist at the edge of magnetic nano-island due to the substrate strain. The strain distribution in our work is also consistent with the previous reports [5].

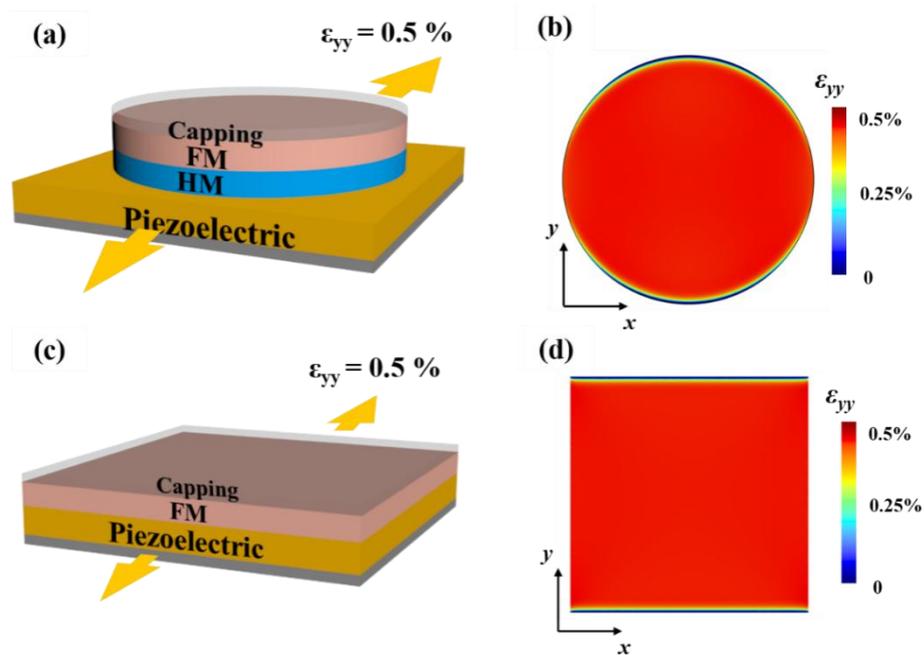

Fig. S1. Schematic diagram of the model of (a) nano-island and (c) nano-film; The distribution of $\varepsilon_{yy}$ in (b) nano-island and (d) nano-film when a uniaxial tensile strain of 0.5% is applied in the y-direction.

**Supplementary Note 2: Energy changes during the application of tensile strain to Néel skyrmion.**

To analyze the variation of the energy in the magnetic domain during the applied strain more deeply, we applied a uniaxial tensile strain of 0.5% to a stable Néel skyrmion to analyze its domain structure and the variation of energy. As shown in Fig. S2(c), we can see that when we apply 0.5% tensile strain in the y-direction, the circular skyrmion will deform and turn into an elongated skyrmion, with the long axis parallel to the direction of the uniaxial strain. Fig. S2(d) shows that the distortion reduces the magnetoelastic energy density and stray field energy density, but will increase the magnetic anisotropy energy density, and we can see that the magnetic anisotropy energy density $F_{anis}$ firstly increases and then decreases to a stable state. At the same time, the stray field energy density $F_{stray}$ firstly decreases and then to a stable state. Therefore, the competition between magnetic anisotropy energy and stray field energy leads to the continuous phase transition and the formation of domain pattern under the uniaxial tensile strain (see Fig. S2(e)).

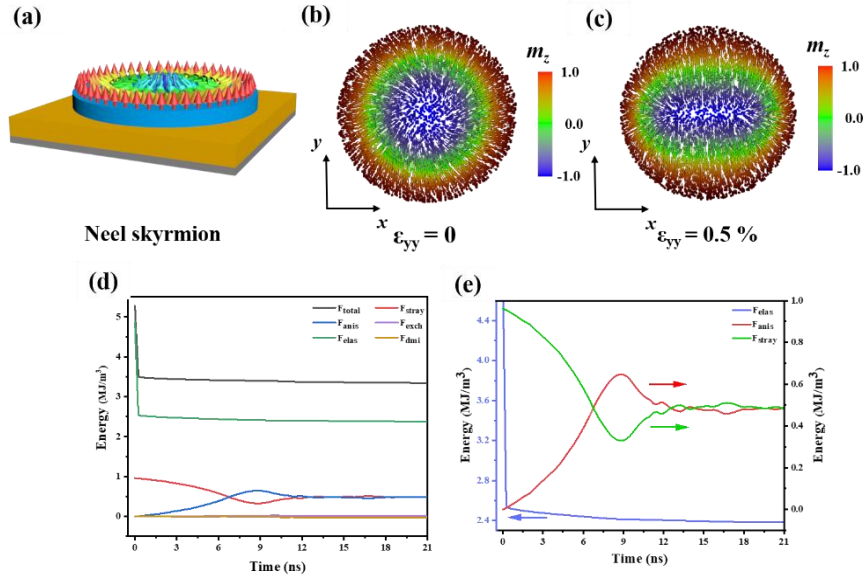

Fig. S2. (a) Schematic diagram of the Néel skyrmion on the nanoisland. The evolution of Néel skyrmion under (b) zero strain and (c) 0.5 % tensile strain in the y-direction. (d) Variations of various energy density terms with time and (e) Variations of magneto-elastic energy density, anisotropy energy density, and stray field energy dendity with time under 0.5% uniaxial tensile strain on Néel skyrmion.

**Supplementary Note 3: Dynamic behavior of the Bloch- and Néel-type magnetic skyrmions under uniaxial tensile strain.**

Fig. S3 shows the dynamic evolution process of the skyrmions with two domain wall types under 0.5% uniaxial tensile strain in the y-direction, respectively. In contrast to the Néel skyrmion, which transforms from a circle to an ellipse [6], the Bloch skyrmion undergoes a process in which the central magnetic moment diminishes and then increases. However, the skyrmion with two different domain wall types eventually exhibits magnetic moments that are parallel to the direction of the uniaxial strain.

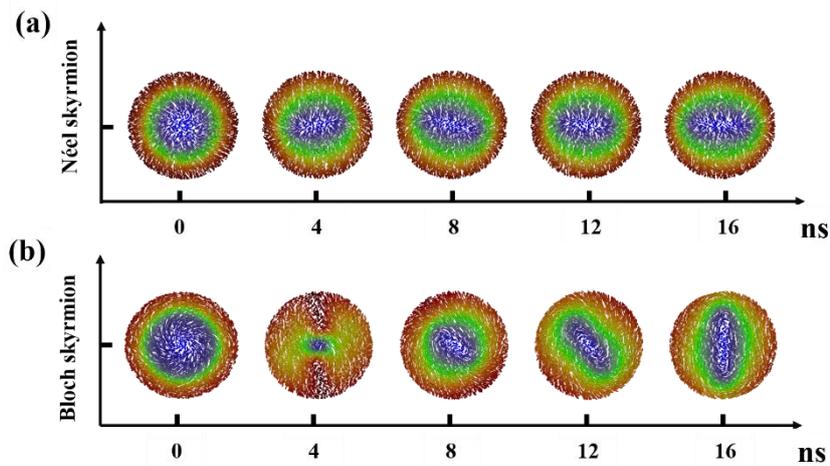

Fig. S3. Dynamic behavior of the (a) Néel skyrmions and (b) Bloch skyrmion under uniaxial tensile strain.